\documentclass[aps,twocolumn,prl,showpacs,showkeys,preprintnumbers,superscriptaddress,nobibnotes,floatfix,nofootinbib]{revtex4-2}

\usepackage{orcidlink}
\usepackage{amsmath}
\usepackage{amsfonts}
\usepackage{amssymb}
\usepackage{mathrsfs}
\usepackage{bm} 
\usepackage{bbm} 
\usepackage{slashed} 
\usepackage{physics} 
\usepackage{esvect} 
\usepackage[version=4]{mhchem}
\usepackage{xcolor} 
\usepackage{graphicx} %
\usepackage{array} 
\usepackage{url} %
\usepackage{amsthm} 
\usepackage{enumerate} 
\usepackage{enumitem} %
\usepackage{booktabs} 
\usepackage{siunitx} 
\usepackage{mdframed} %
\usepackage{tcolorbox} %
\usepackage{fontawesome5} 
\usepackage{braket} 
\usepackage{setspace} %
\usepackage{multirow} %
\usepackage{upgreek} 
\usepackage{xspace} %
\usepackage[normalem]{ulem} 
\usepackage{array}
\usepackage{natbib} 
\usepackage{hyperref} %
\hypersetup{
    colorlinks=true,
    linkcolor=blue,
    citecolor=blue,
    filecolor=magenta,
    urlcolor=blue,
    breaklinks=true
}
\usepackage[all]{hypcap} 
\usepackage[capitalise]{cleveref} %

\setlist[description]{leftmargin=\parindent,labelindent=\parindent} %
\newcommand{\prlsection}[2]{{\it\textbf{#1}{#2}}---}

\usepackage{slashed}

\begin{document}
\hfill TTP26-015

\title{Probing Sub-GeV Dark Matter via Migdal Effect-Induced Electron Excitations}%

\author{Felix Kahlhoefer}
\email{felix.kahlhoefer@kit.edu}
\affiliation{Institute for Astroparticle Physics (IAP), Karlsruhe Institute of Technology (KIT),
Hermann-von-Helmholtz-Platz 1, 76344 Eggenstein-Leopoldshafen, Germany}
\affiliation{Institute for Theoretical Particle Physics (TTP), Karlsruhe Institute of Technology (KIT), 76128 Karlsruhe, Germany}

\author{Liangliang Su}
\email{liangliang.su@kit.edu}
\affiliation{Institute for Astroparticle Physics (IAP), Karlsruhe Institute of Technology (KIT),
Hermann-von-Helmholtz-Platz 1, 76344 Eggenstein-Leopoldshafen, Germany}

\date{\today}

\begin{abstract}
The electron ionization predicted by the Migdal effect in dark matter-nucleus scattering enhances experimental sensitivity to sub-GeV dark matter. In this work, we demonstrate that lower-energy electron excitations provide a novel and promising pathway, enabling the detection of even lighter dark matter particles previously considered inaccessible for direct searches. Direct detection experiments employing a superfluid $^4$He target can exploit this channel by observing electronic excitations via UV-photon emission. We calculate the resulting event rates and find that electron excitations induced by the Migdal effect make it possible to probe dark matter-nucleus scattering for dark matter masses as small as a few MeV.

\end{abstract}

\maketitle

\prlsection{Introduction}{.}
A wealth of astronomical and cosmological observations has established the existence and importance of dark matter (DM) in the Universe. However, no unambiguous evidence for its non-gravitational interactions has been found in laboratory experiments, particularly in searches for weakly interacting massive particles~\cite{Lee:1977ua, Jungman:1995df} via direct detection~\cite{XENON:2023cxc, PandaX:2024qfu, LZ:2024zvo}. Consequently, a growing number of theoretical proposals and novel experiments, such as those based on semiconductors~\cite{SuperCDMS:2018mne, 2022PhRvD.106f2004A, CDEX:2023vvc, SENSEI:2023zdf, DAMIC-M:2025luv} and superfluid helium~\cite{Schutz:2016tid,vonKrosigk:2022vnf,Hirschel:2023sbx,SPICE:2023tru,QUEST-DMC:2023nug}, have begun to explore lighter DM candidates -- so-called sub-GeV DM~\cite{Essig:2011nj,Kouvaris:2016afs,Knapen:2017xzo,Bringmann:2018cvk,Alvey:2019zaa,Wang:2019jtk,Ge:2020yuf,Guo:2020oum,Bell:2021xff,Kahn:2021ttr,Elor:2021swj,Arguelles:2022fqq,Alvey:2022pad,Su:2022wpj,PandaX:2023tfq,Su:2023zgr,Liang:2024xcx,Dutta:2024kuj,Bhattiprolu:2024dmh,Sun:2025gyj,Balan:2025uke,Gong:2025ves,Cheek:2025nul,Li:2025zwg,Ge:2025itf,Bernreuther:2025xqk,Hu:2025dsv,Cox:2025toz,Wang:2026you,Gong:2026dte}. A particularly compelling target for the community is to extend experimental sensitivity down to DM particles as light as 10 MeV, corresponding to the smalles masses for which the freeze-out mechanism is compatible with standard primordial nucleosynthesis~\cite{Sabti:2019mhn,Balan:2024cmq}.

In this context, the Migdal effect has demonstrated remarkable potential for probing sub-GeV DM, as it enables larger detectable energy deposition through electronic ionization and benefits from the lower experimental threshold for electron signals~\cite{Ibe:2017yqa,Dolan:2017xbu,Bell:2019egg,Baxter:2019pnz,Essig:2019xkx,Liang:2019nnx,Liu:2020pat,Knapen:2020aky,Flambaum:2020xxo,Bell:2021zkr,Wang:2021oha,Liang:2022xbu,MIGDAL:2022yip,Blanco:2022pkt,Berghaus:2022pbu,Li:2022acp,Bell:2023uvf,Gu:2023pfg,Herrera:2023xun,He:2024hkr,Kahn:2024nyv,Blanco-Mas:2024ale,Esposito:2025iry,Berghaus:2026kmj}. It arises from the sudden perturbation of the atom induced by DM-nucleus scattering, which induces relative motion between the nucleus and the electron cloud, leading to ionization or excitation of atomic or molecular electrons. As a result, the entire kinetic energy of the DM particle, $m_\chi v_\chi^2 / 2$ with the DM mass $m_\chi$ and velocity $v_\chi$ can potentially be deposited in the detector, while in elastic scattering the maximal momentum transfer is given by $m_\chi v$, corresponding to a nuclear recoil energy of $m_\chi^2 v^2 / 2 m_N$ with the nucleus mass $m_N \gg m_\chi$. Recently, the MARVEL collaboration reported the first direct evidence of the Migdal effect in a neutron bombardment experiment~\cite{Yi:2026fmf}, which further strengthens its viability for sub-GeV DM detection. While most previous studies have focused on ionization signals induced by the Migdal effect, it is worth emphasizing that electronic excitations at lower energy transfer can provide a new and promising channel for probing sub-GeV DM. 

\begin{figure}[t]
    \centering
    \includegraphics[width=0.99\linewidth]{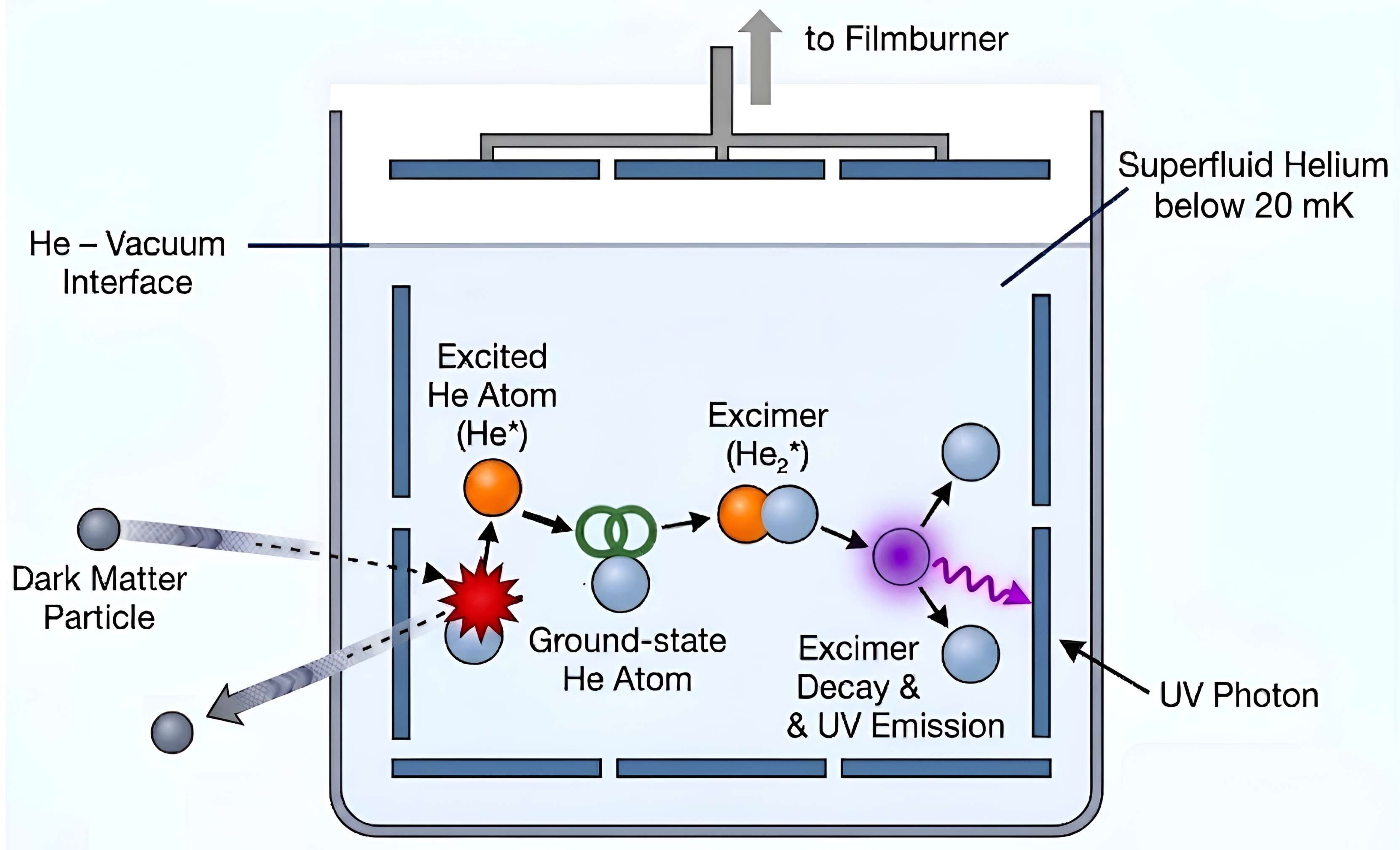}
    \caption{Illustration of a helium excitation process in superfluid $^4$He induced by DM-helium scattering. The DM particle scatters off a helium atom, creating an excitation ($\text{He}^*$) that combines with a ground-state atom to form an excimer ($\text{He}_2^*$). The subsequent radiative decay produces a UV photon, which can be detected by the surrounding magnetic microcalorimeters (MMCs) arrays. The energy absorbed by the recoiling nucleus leads to the evaporation of individual helium atoms, which can be detected by the MMCs in the vacuum phase.}
    \label{fig:process}
\end{figure}

Planned direct detection experiments based on a superfluid $^4$He target and detectors with multiple readout channels, such as DELight~\cite{vonKrosigk:2022vnf} and HeRALD~\cite{SPICE:2023tru}, can potentially probe electron excitations induced by the Migdal effect. In these experiments, an excited helium atom produced by DM-nucleus scattering can combine with another ground-state helium atom to form an excited dimer (an excimer), which subsequently decays via the emission of a ultraviolet (UV) photon, as shown in Fig.~\ref{fig:process}. In the case of elastic DM-nucleus scattering, the excited helium atom arises indirectly when the recoiling nucleus from the DM–helium interaction scatters with surrounding helium atoms, requiring the recoil energy to exceed the excitation threshold ($\sim$20 eV). In contrast, the Migdal effect can directly induce helium excitation from the ground state even when the nuclear recoil energy is well below this threshold, thereby enabling sensitivity to lighter DM particles.

In this Letter, we take the planned DELight experiment as an example to assess the potential of electronic excitation induced by the Migdal effect for sub-GeV DM detection. Assuming fermionic DM interacting with nuclei via a heavy vector mediator, we compute the Migdal-induced excitation probabilities for various helium excitation channels and derive the corresponding event rates. We then obtain, for the first time, sensitivity projections in terms of the DM–nucleon scattering cross section from these electronic excitations in the DELight experiment. Our results demonstrate sensitivity to MeV-scale DM particles and indicate the potential to surpass existing experimental constraints in the mass range of 10--100 MeV.

\prlsection{Electron Excitation induced by the Migdal Effect}{.} 
In the Migdal effect, electron excitation or ionization arises from the sudden relative motion between the recoiling nucleus and the electron cloud. The corresponding transition probability is therefore determined by the overlap between the initial atomic wave function just after the nuclear recoil $|\Psi_{i}^{\prime}\rangle$ and the final state wave function $|\Psi_{f}\rangle$,
\begin{equation}
    \mathcal{P}_{f i} \equiv \left|\left\langle\Psi_{f} \mid \Psi_{i}^{\prime}\right\rangle\right|^{2}=\left|\left\langle\Psi_{f}\right| e^{i \mathbf{q}_{e} \cdot \sum_{k} \mathbf{r}_{k}}| \Psi_{i}\rangle\right|^{2},
\end{equation}
where the second equality relies on the impulse approximation, i.e.\ that the DM–nucleus interaction timescale is short compared to the electronic response time. Consequently, after the collision, the atomic wave function is related to the initial state by a Galilean boost, $|\Psi_{i}^{\prime}\rangle = e^{i \mathbf{q}_{e} \cdot \sum_{k} \mathbf{r}_{k}}|\Psi_{i}\rangle$, where $\mathbf{r}_k$ denotes the position of the $k$-th electron, $\mathbf{q}_e = m_e \mathbf{v}_e = -m_e \mathbf{v}_N$ is the electron momentum determined by the electron mass $m_e$ and the nuclear recoil velocity $\mathbf{v}_N$ and $|\Psi_{i}\rangle$ is the initial atomic wave functions in the rest frame of the nucleus. 

The transition probability of the Migdal effect can be estimated using the dipole approximation at low recoil velocities, where only the active (excited or ionized) electron is considered and multielectron effects are neglected. In this work, we instead describe the electronic structure of the atomic wave function using the multiconfiguration Hartree–Fock method (MCHF)~\cite{grant2007relativistic,2016JPhB...49r2004F}, which incorporates electron correlation effects.\footnote{For a more detailed discussion and a comparison between the dipole approximation and the MCHF method, we refer to Ref.~\cite{Cox:2022ekg}.} In this framework, the atomic state function (ASF) can be expanded as a linear combination of configuration state functions (CSFs) with the same parity $P$, total angular momentum operators, $J$ and $J_z$~\cite{2016JPhB...49r2004F}, 
\begin{equation}
    \Psi_\text {ASF }\left(P J M_{J}\right)=\sum_{\gamma} c_{\gamma} \Phi_{\text {CSFs}}\left(\gamma P J M_{J}\right),
\end{equation}
where the expansion coefficients $c_\gamma$ denote the mixing of different configuration states $\gamma$, and the CSFs can be constructed as the anti-symmetric products of single-electron orbital wave functions $\psi_{n, \kappa, m}$, where ${n, \kappa, m}$ is the principal, Dirac and magnetic quantum number. 

Therefore, the transition probability of the Migdal effect can be rewritten as~\cite{Cox:2022ekg}
\begin{equation}
    \mathcal{P}_{fi}=\bigg|\sum_{\gamma_i, \gamma_f}c_{\gamma_{f}}^{*} c_{\gamma_i} \operatorname{det}\left(M^{\gamma_f\gamma_i}\right)\bigg|^{2},
\label{eq:probability}
\end{equation}
where $c_{\gamma_f}$ and $c_{\gamma_i}$ are the expansion coefficients of the initial and final ASF, respectively. The matrix $M^{\gamma_f\gamma_i}$ is of dimension $N_e \times N_e$ and is constructed from the relativistic single-electron matrix elements $(M^{\gamma_f\gamma_i})_{\beta\alpha} \equiv M_{n \kappa m}^{n^{\prime} \kappa^{\prime} m^{\prime}} \equiv\left\langle\psi_{n^{\prime}, \kappa^{\prime}, m^{\prime}}\right| e^{i \mathbf{q}_{e} \cdot \mathbf{r}}\left|\psi_{n, \kappa, m}\right\rangle\;$.By expanding the phase factor $e^{i \mathbf{q}_{e} \cdot \mathbf{r}}$ in spherical harmonics, we obtain
\begin{equation}
\begin{aligned}
M_{n \kappa m}^{n^{\prime} \kappa^{\prime} m^{\prime}}
 =& \sqrt{4 \pi} \sum_{L, M}(-i)^{L} \sqrt{2 L+1}  Y_{L}^{M}\left(\hat{v}_{N}\right) C_{M}^{L}  \\
& \times \int_{0}^{\infty} d r j_{L}\left(m_{e} v_{N} r\right)\left[P_{n, \kappa}(r) P_{n^{\prime}, \kappa^{\prime}}(r)\right.\\
&\qquad \qquad \qquad \qquad \quad \, \left.+Q_{n, \kappa}(r) Q_{n^{\prime}, \kappa^{\prime}}(r)\right]
\end{aligned}
\end{equation}
where $j_{L}(x)$ is the spherical Bessel function. For a closed-shell initial atom with zero total angular momentum, such as helium or xenon, the transition matrix element is independent of the direction of the recoil velocity. One can therefore choose the $z$-axis along the recoil velocity, yielding $Y_L^{M}(\theta_{\vec{v}}=0, \phi_{\vec{v}}) = \sqrt{(2L+1)/{4\pi}} \delta_{M0}$. 
The angular coefficient $C_{M}^{L}$ includes the Wigner 3-j symbol and some selection rules and is given by  
\begin{equation}
\begin{aligned}
    C_{M}^{L} 
=&(-1)^{2 j^{\prime}-m^{\prime}+1 / 2} \sqrt{(2j+1)(2j^{\prime}+1)} \Pi^{L} _{ll^\prime} \\
   &\times \left(\begin{array}{rrr}
j^{\prime} & L & j \\
-m^{\prime} & M & m
\end{array}\right)\left(\begin{array}{rrr}
j^{\prime} & L & j \\
1 / 2 & 0 & -1 / 2
\end{array}\right),
\end{aligned}
\label{eq:C_LM}
\end{equation}
here the parity selection rule $\Pi^{L} _{ll^\prime}$ equals unity if $l+l^{\prime} + L$ is even and zero otherwise, with $l(l^\prime)$ denoting the orbital angular momentum quantum numbers of the initial (final) electron. 

The functions $P(r)$ and $Q(r)$ are the large and small radial components of the relativistic electron wave function $\psi_{n, \kappa, m}$, respectively. In this work, we employ the \texttt{GRASP2018} package~\cite{FROESEFISCHER2019184, atoms11040068} to compute the electron radial wave functions for both the ground and excited states of the helium atom, as well as the expansion coefficients of the CSFs. For the ground state, only the $1s^2$ configuration is included, with no electronic excitation. For the excited states, the radial wave functions are obtained from configuration expansions including single-electron excitations from the outer shell, implemented iteratively by gradually increasing the maximum principal quantum number $n$ while keeping lower-$n$ orbital wave functions fixed. 

\begin{figure}
    \centering
    \includegraphics[width=0.99\linewidth]{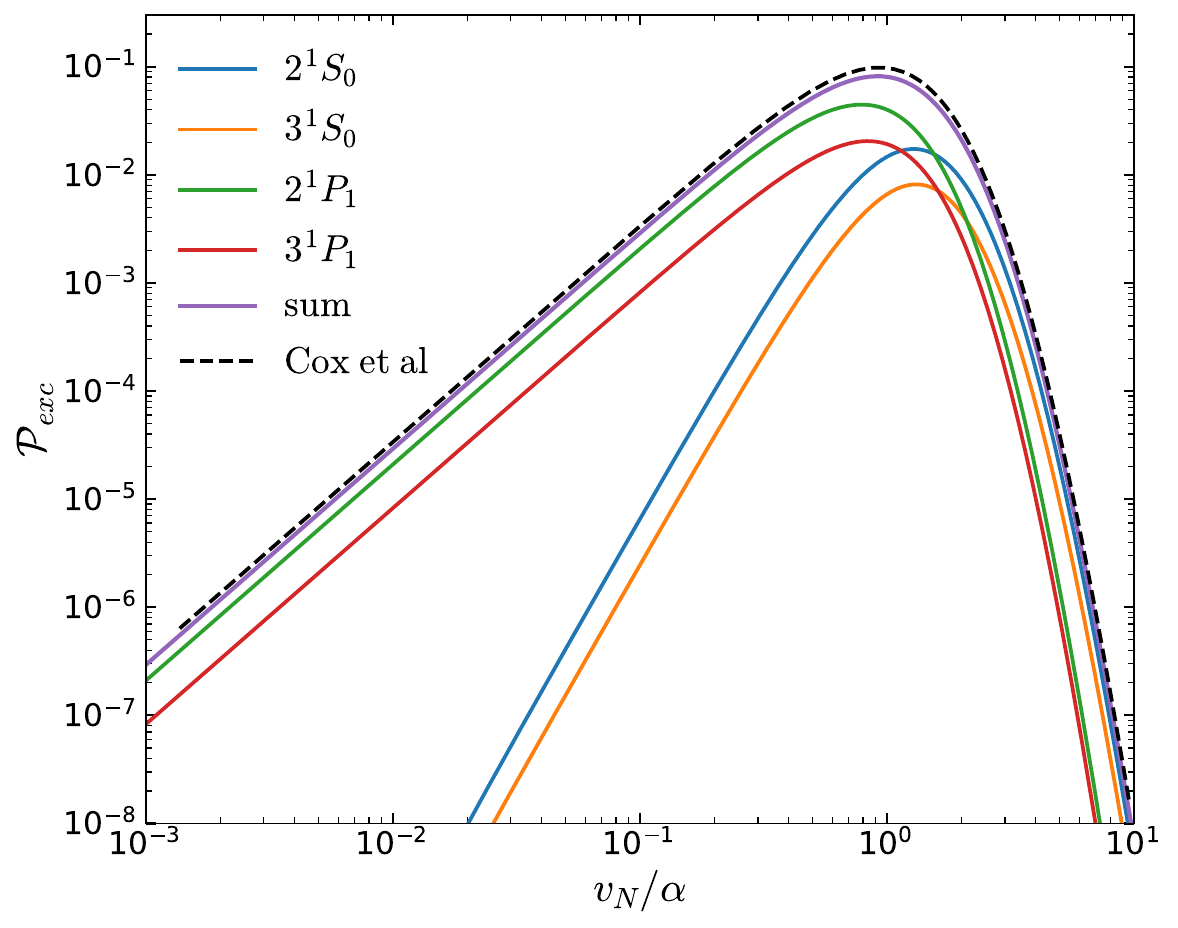}
    \caption{The electron excitation probabilities of the Migdal effect in neutral particle-helium scattering as a function of the nuclear recoil velocity $v_N$. The solid lines show transitions from the ground state to $2 ^1S_0$ (blue), $2 ^1P_1$ (green), $3 ^1S_0$ (orange), and $3 ^1P_1$ (red), as well as their sum (purple). The black dashed line denotes the result from Ref.~\cite{Cox:2022ekg}. }
    \label{fig:exc_probability}
\end{figure}

For our analysis, we only consider the dominant excited states of the helium atom with principal quantum number $n\leq 3$. In Fig.~\ref{fig:exc_probability}, we show the electron excitation probabilities of helium induced by the Migdal effect as a function of nuclear velocity $v_N$, expressed in units of the fine-structure constant $\alpha$, for transitions from the ground state to excited states $2 ^1S_0$, $2 ^1P_1$, $3 ^1S_0$, and $3 ^1P_1$, respectively. The dominant excited states only include the singlet excited states, because the Wigner 3j-symbol in Eq.~\ref{eq:C_LM} imposes the selection rule $-m^{\prime}+M+m =0$ for the magnetic quantum numbers of the initial and final states. In other words, the phase factor $e^{i \mathbf{q}_{e} \cdot \mathbf{r}}$ induced by the Migdal effect corresponds to a dipole-like (vector) interaction operator that conserves the total spin of the helium atom.
We find that the dominant excitation channels depend on the nuclear recoil velocity $v_N$. In the high-velocity regime, $v_N \gtrsim \alpha$, the excitation to the $2 ^1S_0$(blue line) and $3 ^1S_0$ (orange line) states dominates. In contrast, for lower velocities, $v_N \lesssim \alpha$, the contributions from the $2 ^1P_1$  (green line) and $3 ^1P_1$ (red line) states become more significant. The latter region corresponds to the typical kinematics of the Migdal effect induced by sub-GeV DM in the Milky Way halo.
Finally, we show in Fig.~\ref{fig:exc_probability} a comparison of the sum of the considered states (purple solid line) with the results of Ref.~\cite{Cox:2022ekg} (black dashed line) and find good agreement. The small residual difference arises from contributions of higher excited states. 

\prlsection{Calculation of differential event rates}{.} 
In this work, we consider standard spin-independent DM scattering, which could arise for example from the interactions of a fermionic DM particle with quarks via the exchange of a heavy mediator. The differential event rate of electron excitation induced by the Migdal effect in DM-helium scattering is then given by 
\begin{equation}
    \frac{\mathrm{d}^2 R}{\mathrm{~d} E_{R} \mathrm{~d} E_{\mathrm{EM}}} = \frac{\mathrm{d} R_N}{\mathrm{d} E_R} \sum_i \mathcal{P}^{(i)}_{e x c}\left(E_{R}\right) \delta(E_{\mathrm{EM}}-E_{\mathrm{exc}}^{(i)})  \label{eq:d2R} \; .
\end{equation}
where the differential event rate of the elastic DM-nucleus scattering $\mathrm{d}R_N/\mathrm{d} E_R$ can be found in the Appendix. The sum in eq.~\eqref{eq:d2R} runs over the different excitation states included in the calculation. the corresponding excitation energies $E_{\mathrm{exc}^{(i)}}$ of the $^4$He atom are taken from the NIST database~\cite{NIST_ASD} and shown in Tab.~\ref {tab:exc_energy}. The corresponding electron excitation probabilities induced by the Migdal effect, $\mathcal{P}^{(i)}_{\mathrm{exc}}$, depend on the nuclear recoil velocity (or energy $E_R$), with $v_N = \sqrt{2E_R/m_{\mathrm{He}}}$, where $m_{\mathrm{He}}$ is the mass of the $^4$He atom. Finally, the Dirac delta function $\delta(E_{\mathrm{EM}}-E_{\mathrm{exc}^{(i)}})$ ensures energy conservation.

\begin{table}[!ht]
\centering
\begin{tabular}{c c c c c}
\toprule
State & $2\,^1{S}_0$ & $2\,^1{P}_1$ & $3\,^1{S}_0$ & $3\,^1{P}_1$ \\
\midrule
$E_{\mathrm{exc}}$ [eV] & 20.616 & 21.218 & 22.920 & 23.087 \\
\bottomrule
\end{tabular}
\caption{Excitation energies of the $^4$He atomic  excited states~\cite{NIST_ASD}.}
\label{tab:exc_energy}
\end{table}

\begin{figure}
    \centering
    \includegraphics[width=0.95\linewidth]{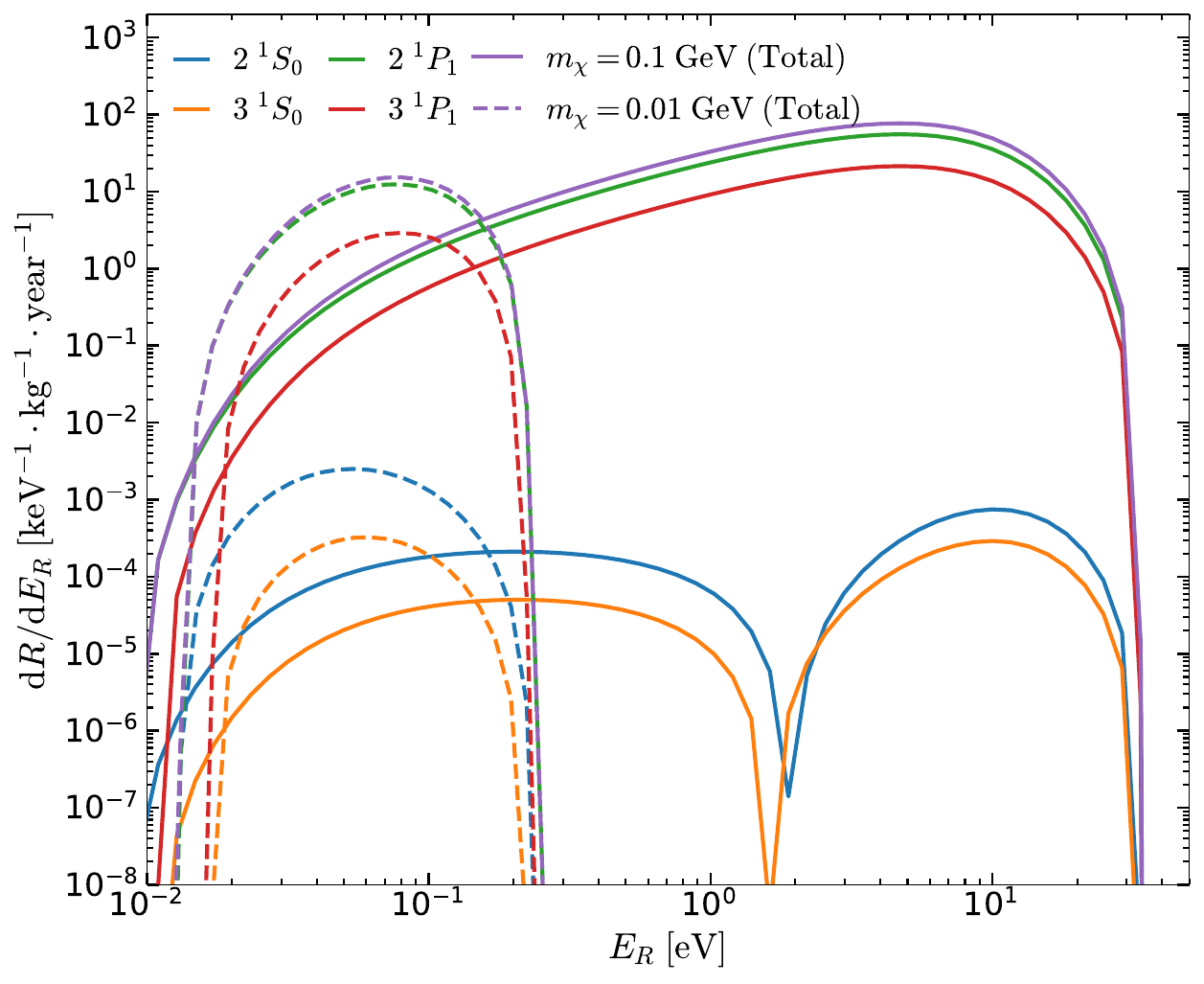}
    \caption{The differential event rate $\mathrm{d}R/\mathrm{d} E_R$ as a function of nuclear recoil energy $E_R$ for $2 ^1S_0$ (blue lines), $2 ^1P_1$ (green lines), $3 ^1S_0$ (orange lines), and $3 ^1P_1$ (red lines) of helium induced by the Migdal effect, and their sum (purple lines). The solid and dashed lines correspond to different DM masses, $m_\chi =0.1$ and 0.01 GeV, respectively. Here we take the cross section $\sigma^{\mathrm{SI}}_{\chi n} = 10^{-38}\;\mathrm{cm}^2$ as the benchmark. }
    \label{fig:dRdER}
\end{figure}

In Fig.~\ref{fig:dRdER}, we show the differential event rate $\mathrm{d}R/\mathrm{d}E_R$ for each of the four excitation processes induced by the Migdal effect as well as their sum, assuming $m_\chi = 0.01\;\mathrm{GeV}$ (dashed lines) and $0.1\;\mathrm{GeV}$ (solid lines) with a reference spin-independent DM–nucleon scattering cross section $\sigma^{\mathrm{SI}}_{\chi n} = 10^{-38}\;\mathrm{cm}^2$. As discussed above, the excitations from the ground state to the $2 ^1P_1$ state (green lines) dominate for sub-GeV DM.  Unsurprisingly, the heavier DM particle (with larger kinetic energy) can deposit larger recoil energy. In contrast to elastic scattering, however, we find that the differential event rate peaks for finite recoil energy. This is because $v_\text{min}$ in eq.~\eqref{eq:vmin} of the Appenix diverges for $E_\text{EM} > 0$ and $E_R \to 0$. Since scattering is possible only if $v_\text{min}$ is smaller than the maximal DM velocity in the halo, i.e.\ if $v_{\min} < v_{\mathrm{esc}} + |\vec{v}_E|$, the nuclear recoil energy cannot be arbitrarily small. An important implication is that in superfluid helium the electron excitation is always accompanied by a second signal that can be detected experimentally together with the UV photon.

For $m_\chi = 0.1\;\mathrm{GeV}$, the differential rates for the $2 ^1S_0$ and $3 ^1S_0$ states exhibit a peculiar feature around $E_R \simeq 2\;\mathrm{eV}$. This behavior originates from a slight loss of orthogonality between the initial and final electron wave functions in the numerical implementation of the \texttt{GRASP2018} package, i.e., $\langle \psi_{2s} | \psi_{1s} \rangle$ is not exactly zero. This leads to a non-zero value of the integral $\int_{0}^{\infty} d r j_{L}\left(m_{e} v_{N} r\right)\left[P_{n, \kappa}(r) P_{n^{\prime}, \kappa^{\prime}}(r)+Q_{n, \kappa}(r) Q_{n^{\prime}, \kappa^{\prime}}(r)\right]$ in the limit $v_N\to 0$ for the dominant $L=0$ term of the $2^1S_0$ and $3^1S_0$ states, causing an overestimation of the contributions at low recoil velocity (energy). Ref.~\cite{Flambaum:2020xxo} proposes a “subtract-one” prescription, replacing the radial integral $\int_{0}^{\infty} \mathrm{d}r j_{0}\left(m_{e} v_{N} r\right)\cdots$ with $\int_{0}^{\infty} \mathrm{d}r \left[j_{0}\left(m_{e} v_{N} r\right) - 1\right]\cdots$, which resolves the issue. In any case, this issue does not affect the $L=1$ contributions from the $2^1P_1$ and $3^1P_1$ states, which give by far the dominant contribution to our final results. 

\prlsection{Sensitivity projections}{.} 
To estimate the potential relevance of electron excitation induced by the Migdal effect, we consider the planned superfluid $^4$He experiment DELight as the benchmark detector. The DELight experiment relies on the conversion of the deposited nuclear recoil energy into multiple quasiparticle excitations, including phonons, rotons, scintillation photons, and metastable excimer states, as shown in Fig.~1 of Ref.~\cite{DELight:2024bgv}. These excitations can be efficiently collected and measured by highly sensitive cryogenic sensors, such as MMCs, enabling a precise calorimetric readout with very low energy thresholds. In particular, excimer states produced in the electron excitation channel can be detected via their UV photon emission. Ordinary nuclear recoil processes generate two types of excited states, namely singlet and triplet states, which subsequently combine with a ground-state helium atom to form the corresponding excimers, $\mathrm{A}^1\Sigma_u^+$ (singlet) and $\mathrm{a}^3\Sigma_u^+$ (triplet). Since the triplet states are very long-lived, they are impossible to associate with the primary recoil event and effectively constitute an energy loss channel. Fortunatley, for electron excitations via the Migdal effect, only singlet excimers are formed, since the total spin of the helium atom is conserved. In principle, there can still be some reduction of the effective UV photon yield through non-radiative channels such as Penning quenching at high excitation densities and impurity-induced losses. However,  since singlet excimers induced by the Migdal effect are rare and have a lifetime of a few nanoseconds, we assume a 100\% radiative decay probability, i.e.\ a UV photon fraction of unity.

In addition, the recoil energy of the helium nucleus is converted into quasiparticles (phonons and rotons), which propagate through the superfluid helium and bounce of the walls until they reach the helium-vacuum interface and cause the evaporation of individual helium atoms. These atoms can be absorbed and detected by MMCs in the vacuum phase, providing a second signal in coincidence with the UV photon.

\begin{figure}
    \centering
    \includegraphics[width=0.95\linewidth]{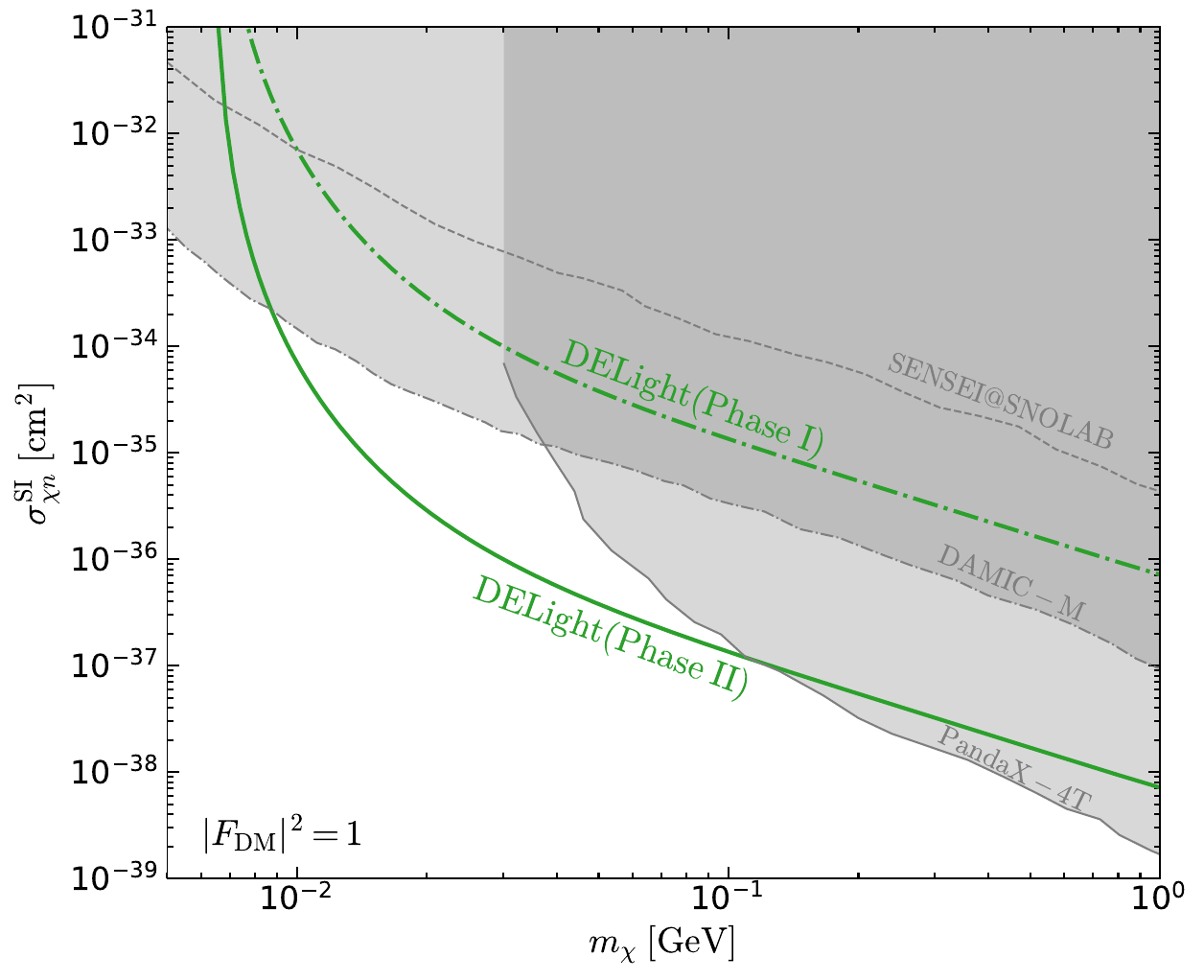}
    \caption{The projected sensitivity for Migdal-induced electron excitation signals in the DELight experiment at 90\% C.L. in terms of the DM–nucleon scattering cross section $\sigma_{\chi n}^{\mathrm{SI}}$. The green dashed-dotted line corresponds to Phase I (1 kg-day exposure), while the green solid line corresponds to Phase II (100 kg-day exposure). The gray shaded region indicates existing constraints from Migdal-induced ionization signals in the PandaX-4T~\cite{PandaX:2023xgl}, SENSEI~\cite{SENSEI:2023zdf}, and DAMIC-M experiments~\cite{DAMIC-M:2025luv}.}
    \label{fig:limit}
\end{figure}

The current plan for the DELight experiment consists of two phases: Phase I will operate a 1-litre superfluid $^4$He target in order to achieve a total exposure of 1 kg$\cdot$day, while Phase II will scale up to a 10-litre detector aiming for a total exposure of 100 kg$\cdot$day~\cite{Solmaz:2025xiq}. While the precise detector efficiency and energy threshold are not yet known, we assume that events containing a UV photon can always be detected.  If backgrounds for such processes are negligible, the projected sensitivity at 90\% C.L. includes all parameter points that predict at least 2.3 events. The resulting sensitivity projections are shown in Fig.~\ref{fig:limit}. We conclude that the electron excitations induced by the Migdal effect can probe DM masses down to less than 10 MeV, thus covering the entire mass range relevant for thermal sub-GeV DM. Although the Phase I result (green dashed-dotted line) is weaker than the Migdal effect-based constraints from the DAMIC-M experiment (gray dashed-dotted line)~\cite{DAMIC-M:2025luv}, the Phase II sensitivity (green solid line) surpasses existing Migdal effect-induced limits from other experiments in the 10–100 MeV mass range. 

{To conclude our analysis, we emphasize that our results are based on the Migdal effect computed for isolated helium atoms, even though helium in the DELight experiment is in the superfluid phase. Specifically, we assume that we can treat the primary interaction of DM as scattering off individual nuclei and neglect collective excitations, and that we can calculate transition probabilities using the electron wave functions of free helium atoms. The first assumption is expected to be valid, because the de Broglie wavelength associated with nuclear recoils of energy $E_R \gtrsim 0.01\;\mathrm{eV}$ (see Fig.~\ref{fig:dRdER}), $\lambda \lesssim 2\pi/\sqrt{2 m_{\mathrm{He}} E_R} \sim 1.4\;\mathrm{\AA}$, which is smaller than the average interatomic spacing in superfluid helium $d_{\mathrm{He}} \sim (\rho_{\mathrm{He}}/m_{\mathrm{He}})^{-1/3}\sim 3.6\;\mathrm{\AA}$, where $\rho_{\mathrm{He}} = 0.145 \;\mathrm{g}/\mathrm{cm}^{3}$ is the density of superfluid helium~\cite{10.1063/1.556028}. To validate the section assumption, we note that the contribution of orbital electron wave functions are concentrated within their root mean square radii, with $\sqrt{\langle r^2\rangle} \sim 0.58\;\mathrm{\AA}$ for the ground state~\cite{fischer1977hartree} and $\sim 2.9\;\mathrm{\AA}$ for the $2p$ orbital associated with the dominant $2^1P_1$ state~\cite{jcp1s2p}, both of which remain smaller than the interatomic spacing. Therefore,  the isolated helium atoms assumption can provide a good approximation for estimating the Migdal effect in superfluid helium.}

\prlsection{Conclusion}{.} 
In this work, we have demonstrated that Migdal-induced electron excitations provide a novel and sensitive channel for sub-GeV dark matter detection, outperforming the ionization channel in certain regions of parameter space. Using relativistic atomic wave function calculations for helium, we compute excitation probabilities and identify the dominant singlet excimer channels relevant for the Migdal effect. Applying these results to the upcoming DELight superfluid $^4$He experiment, we show that UV photons from excimer decay enable efficient detection of electron excitations induced by the Migdal effect. Assuming a zero-background scenario, we derive projected 90\% C.L. sensitivities for two experimental phases, with Phase II reaching sensitivity to DM masses at the MeV scale and potentially improving existing constraints based on the Migdal effect in the 10--100 MeV range by more than an order of magnitude. These results establish electron excitation induced by the Migdal effect as a powerful complementary probe for sub-GeV DM in superfluid helium detectors. This process may in principle be extended to liquid xenon and argon detectors and could lead to competitive or improved sensitivities once low-energy backgrounds are under control.

\section*{Acknowledgements}
We thank Rahel Gabriel for collaboration in early stages of this work, Peter Cox, Matthew Dolan and Christopher McCabe for many valuable discussion on the transition probability calculations and the members of the DELight Collaboration for discussions. The initial draft of Fig.~1 has been generated with the assistance of Gemini. This work has been supported by the Alexander von Humboldt Foundation and the Heidelberg Karlsruhe Strategic Partnership (HEiKA). This work was supported by the Munich Institute for Astro-, Particle and BioPhysics (MIAPbP), which is funded by the Deutsche Forschungsgemeinschaft under Germany´s Excellence Strategy -- EXC--2094 -- 390783311.

\section{Appendix}
Assuming that DM particles interact with the nucleus via a heavy vector mediator, the differential event rate for elastic DM–nucleus scattering can be written as follows
\begin{align}
\frac{\mathrm{d} R_N}{\mathrm{~d} E_{R} } = & \frac{\rho_{\chi} \sigma_{\chi n}^{\mathrm{SI}}}{2 m_{\chi} \mu_{\chi n}^{2}}[f_p Z + f_n (A-Z)]^2 \nonumber \\ & \times \left|F_{\mathrm{N}}\left(E_{R}\right)\right|^{2} \left|F_{\mathrm{DM}}\left(E_{R}\right)\right|^{2} \eta_{\chi}\left(v_{\min }\right)  \; .
\label{eq:d2RN}
\end{align}
Here, $\rho_\chi =0.3 \;\mathrm{GeV}/\mathrm{cm}^3$ is the local DM density, $m_\chi$ and $\mu_{\chi n}$ denote the DM mass and the reduced mass of the DM-nucleon system, respectively, and $\sigma_{\chi n}^{\mathrm{SI}}$ is the spin-independent DM–nucleon scattering cross section at zero momentum transfer. For the $^4$He atom, $A=4$ and $Z=2$ are the atomic mass and charge number, respectively, and the parameters $f_{p,n}$ are dimensionless DM couplings to the proton and neutron, which we set to $f_p =f_n =1$. 

In the second line of eq.~\eqref{eq:d2RN}, the nuclear form factor $|F_N|^2$ accounts for the finite-size structure of the helium nucleus. For light nuclei such as helium, a dipole form factor is commonly adopted, $|F_N(Q^2)|^2 = 1/\left(1 + Q^2/\Lambda_{\mathrm{He}}^2\right)^4$~\cite{Perdrisat:2006hj}, where $Q^2 = 2 m_{\mathrm{He}} E_R$ is the squared momentum transfer and $\Lambda_{\mathrm{He}} = 0.41\,\mathrm{GeV}$ characterizes the nuclear size~\cite{ANGELI2004185}. The DM form factor $F_\text{DM}$, which includes additional properties of the interaction, is set to unity for a heavy mediator in our analysis.

The inverse mean speed function includes the information on the DM velocity distribution, and is defined as 
\begin{equation}
\eta\left(v_{\min }\right)=\int \mathrm{d}^{3} \vec{v}_{\chi} f\left(\vec{v}_{\chi}\right) \frac{1}{v_{\chi}} \Theta\left(v_{\chi}-v_{\min }\right),
\end{equation}
where $f(\vec{v}_{\chi})$ denotes the velocity distribution of DM with velocity $\vec{v}_{\chi}$. Neglecting the time dependence in the Earth's frame, the velocity distribution can be written as
\begin{equation}
    f\left(\vec{v}_{\chi}\right)=\frac{1}{K} \exp(-\frac{\left|\vec{v}_{\chi}+\vec{v}_{E}\right|^{2}}{v_{0}^{2}}) \Theta\left(v_{\mathrm{esc}}-\left|\vec{v}_{\chi}+\vec{v}_{E}\right|\right),
\end{equation}
where we adopt $v_0 = 220\;\mathrm{km/s}$ as the characteristic velocity, $v_{\mathrm{esc}} = 544\;\mathrm{km/s}$ as the Galactic escape velocity, and $|\vec{v}_E| = 232\;\mathrm{km/s}$ as the average velocity of the Earth relative to the DM halo~\cite{Radick:2020qip}. The parameter $K$ denotes the normalization factor and is given by $K = v_0^3 \pi \left[\sqrt{\pi}\;\mathrm{Erf}\!\left(\dfrac{v_{\mathrm{esc}}}{v_0}\right) - 2 \dfrac{v_{\mathrm{esc}}}{v_0} \exp\!\left(-{v_{\mathrm{esc}}^2}/{v_0^2}\right)\right]$.

The minimum velocity that a DM particle must have in order to scatter inelastically and induce a nuclear recoil with energy $E_R$ and an electromagnetic energy $E_{\mathrm{EM}}$ is given by
\begin{equation}
    v_{\min }=\sqrt{\frac{m_{\mathrm{He}} E_{\mathrm{R}}}{2 \mu^{2}_{\chi \mathrm{He}}}}+\frac{E_{\mathrm{EM}}}{\sqrt{2 m_{\mathrm{He}} E_{\mathrm{R}}}} \; , \label{eq:vmin}
\end{equation}
where $\mu_{\chi \mathrm{He}}$ is the reduced mass of the DM particle and $^4$He atom.

\bibliography{refs}

\end{document}